# Structural versus Allocative Inefficiency Across Organizational Settings

**Author:** Jan van de Poll

**Affiliation:** Transparency Lab Research Institute

**Abstract**

Organizations invest heavily in internal mechanisms such as incentive systems, governance structures, performance measurement, analytics, and repeated reorganizations. Yet realized performance gains are often weak or short-lived, particularly in large and mature organizations. Standard explanations emphasize allocative inefficiency arising from incentive misalignment, information problems, or bounded rationality. While relevant, these explanations struggle to account for environments in which effort is high, optimization capabilities are sophisticated, and aggregate performance remains far below apparent potential.

This paper develops a theory of structural inefficiency inside organizations. The central mechanism is the presence of internal externalities: unpriced cross-effects of organizational mechanisms acting simultaneously on multiple performance objectives. When such effects are mixed in sign, organizational effort is partially canceled internally, creating a structural ceiling on realized performance. The paper introduces a simple representation of organizational primitives and key performance indicators and derives a realized-performance law in which internal conflict enters as a multiplicative discount factor. Under weak and realistic conditions, structural inefficiency dominates allocative inefficiency, explaining why optimization often fails before organizational coherence is restored.

## 1. Introduction

Organizations are complex systems designed to pursue multiple objectives simultaneously. Modern firms rarely optimize a single performance dimension; instead, they balance cost efficiency, quality, speed, compliance, innovation, safety, customer satisfaction, and employee retention, among others. To manage these objectives, organizations deploy a growing array of internal mechanisms: incentive schemes, governance rules, approval processes, reporting requirements, information systems, performance metrics, and formal authority structures. These mechanisms are continually revised and expanded in response to perceived performance gaps, environmental change, regulatory pressure, or strategic ambition.

A striking empirical regularity accompanies this process. Despite sustained investment in organizational improvement, realized performance gains are often weak, unstable, or disappointing relative to expectations. Organizations become busier, more measured, more controlled, and more "aligned," yet net outcomes improve slowly, if at all. Gains along one dimension are frequently offset by losses along another. New initiatives create activity and visibility but fail to generate durable improvement.

This pattern seems neither sector-specific nor limited to poorly managed firms. It appears in private corporations, public agencies, and non-profit organizations; in competitive markets and regulated environments; and in firms widely regarded as professionally managed. Moreover, it persists even as organizations adopt increasingly sophisticated optimization tools, including advanced analytics, performance dashboards, and algorithmic decision support.

### Standard explanations and their limits

The dominant explanations for organizational underperformance emphasize allocative inefficiency. In economic terms, organizations fail because effort is misallocated: incentives are imperfectly aligned, information is incomplete, decision rights are poorly assigned, or managers are boundedly rational (Jensen & Meckling, 1976; Simon, 1955). A large literature examines how better contracts, stronger incentives, improved information, or superior search processes can improve outcomes.

Closely related is the literature on multitasking and performance measurement, which shows that rewarding observable tasks can distort behavior toward measured dimensions at the expense of unmeasured ones (Ridgway, 1956; Holmström and Milgrom, 1991; Courty and Marschke, 2008). From this perspective, poor outcomes arise because organizations choose the wrong metrics or weight them incorrectly.

A third strand focuses on organizational design and coordination, documenting persistent conflict and ambiguity in complex structures such as matrix organizations (Kolodny, 1979; Larson and Gobeli, 1987). Here, inefficiency is attributed to unclear authority, overlapping responsibilities, or coordination costs.

While each of these perspectives captures important aspects of organizational reality, they leave a significant residual puzzle. In many organizations:

- effort is high rather than low,
- incentives are strong rather than weak,
- measurement is extensive rather than sparse,
- optimization capabilities are improving rather than deteriorating.

Yet realized performance remains structurally constrained. Improvements in one domain often require compensating sacrifices elsewhere, producing a sense of stagnation despite continuous reform.

**A recurring empirical pattern: organizational layering**

A key empirical regularity cuts across these literatures: organizational change is overwhelmingly additive. When problems recur, organizations rarely remove existing mechanisms. Instead, they add new ones.

When quality problems arise, firms add quality controls. When compliance failures occur, they add compliance layers. When coordination becomes difficult, they add reporting lines, committees, or matrix dimensions. When performance is unclear, they add metrics and dashboards. Each intervention is typically justified in isolation and often improves the targeted objective. Yet older mechanisms remain in place, and the organization gradually accumulates layers of objectives, controls, and accountabilities.

This pattern is particularly visible in what practitioners colloquially describe as “cheese-grater” organizations: organizations that have been repeatedly reorganized by slicing authority and accountability along new dimensions—product, function, region, customer, risk, ESG, or project—without removing previous ones. Over time, the same roles, processes, and systems are required to satisfy an expanding set of partially conflicting demands.

Importantly, cheese-grater organizations are not accidents or failures of intent. They are typically the outcome of rational attempts to internalize trade-offs and represent multiple perspectives simultaneously. Each added dimension addresses a real concern. The puzzle is not why organizations adopt such structures, but why the cumulative effect so often appears to reduce rather than enhance net performance.

### Structural inefficiency and internal conflict

This paper argues that the common element underlying these phenomena is structural inefficiency arising from internal conflict among organizational mechanisms. The core idea is simple but under-formalized: internal mechanisms often affect multiple performance objectives simultaneously, and those effects are frequently mixed in sign. When many such mechanisms operate together, organizational effort is partially canceled internally.

This cancellation is not a consequence of poor execution or irrational behavior. It is a structural property of how mechanisms interact. An organization may exert substantial effort, deploy high-leverage mechanisms, and allocate resources competently, yet realize only a fraction of the potential impact because positive effects on some objectives are offset by negative effects on others.

Crucially, this form of inefficiency is distinct from misallocation. Allocative inefficiency concerns where effort is applied. Structural inefficiency concerns what fraction of effort can ever translate into net improvement given the organization’s internal configuration.

### Contribution and structure of the paper

The paper makes three contributions.

First, it develops a formal framework that distinguishes magnitude, direction, and conflict in organizational mechanisms and shows how conflict enters realized performance as a structural discount. This makes internal conflict analytically tractable and separable from effort allocation.

Second, it applies this framework across several organizational settings that are usually studied in isolation— incentive-intensive systems, governance-heavy organizations, KPI-saturated performance regimes, and cheese-grater organizational forms—showing that they generate internal conflict through the same underlying mechanism.

Third, it demonstrates that under weak and empirically plausible conditions, structural inefficiency dominates allocative inefficiency. This explains why optimization efforts often fail until structural coherence is restored and why repeated organizational reform can generate activity without commensurate improvement.

The remainder of the paper proceeds as follows. Section 2 introduces the formal framework. Sections 3–7 apply the framework to different organizational settings, including an integrated treatment of governance layering, KPI proliferation, and cheese-grater reorganization. Subsequent sections synthesize the results and derive implications for optimization and organizational change.

## 2. A structural framework for organizational inefficiency

This section develops a minimal formal framework for analyzing internal organizational inefficiency. The objective is not to model any particular organizational form, but to isolate a general structural mechanism that operates across settings. The framework is intentionally sparse: it introduces only those objects needed to distinguish effort, leverage, and internal conflict, and to separate allocative inefficiency from structural inefficiency in a precise way.

The central idea is that organizational mechanisms often affect multiple performance objectives simultaneously. When those effects are mixed in sign, organizational effort can cancel internally. This cancellation is not a behavioral anomaly; it is a structural property of how mechanisms interact with objectives. The framework below formalizes this intuition.

**Organizational primitives and performance objectives**

Consider an organization that pursues a set of performance objectives indexed by

$$k = 1, \dots, K.$$

These objectives are represented by key performance indicators (KPIs), which may include cost efficiency, speed, quality, compliance, safety, innovation, customer satisfaction, employee retention, or other relevant outcomes. KPIs are normalized so that higher values correspond to better performance. The framework does not assume that KPIs are perfectly measured or complete; it assumes only that they represent the dimensions along which organizational performance is evaluated and acted upon.

The organization influences these objectives through a set of organizational primitives, indexed by

$$i = 1, \dots, N.$$

A primitive is defined as a concrete, recurring, and verifiable organizational mechanism or behavior that can be directly observed or audited—such as the existence of formal one-on-one meetings (in HRM) or mandatory approval steps and centrally maintained risk logs (in ICT), or the approach to greeting customers and fitting room cleanliness (in fashion retail)—rather than an abstract norm or perception. Primitives persist over time and shape behavior repeatedly.

Two features of primitives are important. First, primitives are structural: they shape behavior repeatedly, not just in a single decision. Second, primitives are multi-objective: most primitives affect more than one KPI. A control may reduce risk but slow execution; an incentive may increase output but reduce quality; a reporting requirement may improve transparency but increase administrative burden.

We capture these effects by defining, for each primitive $i$ and KPI $k$, a signed effect:

$$b_{ik} \in \mathbb{R}.$$

A positive value indicates that the primitive improves the KPI, a negative value indicates deterioration, and zero indicates no material effect. The collection of effects $\{b_{ik}\}$ represents the organization's internal causal structure.

Importantly, the framework does not require that all effects be known with precision. It requires only that effects correspond, in principle, to observable organizational consequences rather than subjective opinions. This assumption is consistent with the literature on performance measurement and organizational analysis, which emphasizes observable behavioral and outcome-based indicators (Ridgway, 1956; Meyer and Gupta, 1994).

**Magnitude and direction: separating leverage from alignment**

When a primitive affects only a single KPI, its evaluation is straightforward. When it affects many KPIs, evaluation becomes more complex because improvements along one dimension may be offset by deteriorations along another. To disentangle leverage from alignment, we introduce two summary measures for each primitive.

**Magnitude**

Define the **magnitude** of primitive $i$ as:

$$M_i = \sum_{k=1}^{K} | b_{ik} |$$

Magnitude captures the total strength of the primitive's influence across all objectives, irrespective of sign. A primitive with large positive and negative effects across many KPIs has high magnitude. Intuitively, magnitude measures how much the organization "moves" when effort is applied to the primitive.

Magnitude is closely related to managerial salience. High-magnitude primitives tend to attract attention, resources, and effort because they appear to offer substantial leverage. Governance systems, incentive schemes, and major IT platforms often fall into this category precisely because they affect many outcomes simultaneously.

**Net direction**

Define the **net direction** of primitive $i$ as:

$$D_i = \sum_{k=1}^{K} b_{ik}$$

Net direction aggregates the signed effects and captures whether the primitive is beneficial or harmful on balance. A primitive may have large magnitude but small net direction if its positive and negative effects largely offset.

Separating magnitude from direction is essential. Many organizational mechanisms are powerful in the sense that they affect many outcomes, yet controversial or ambiguous in their net contribution. This ambiguity is often treated informally in organizational discourse; here it is formalized.

**Internal conflict as a structural property**

Magnitude and direction alone are insufficient to characterize how a primitive translates effort into performance. What matters is the degree to which its effects reinforce or cancel one another across objectives.

Define **internal conflict** for primitive $i$ as:

$$C_i = 1 - \frac{|D_i|}{M_i}, C_i \in [0,1]$$

Conflict measures the fraction of a primitive's total influence that is lost to cross-objective cancellation.

- If all effects have the same sign, then $|D_i| = M_i$ and $C_i = 0$. The primitive is perfectly aligned.
- If positive and negative effects largely offset, then $|D_i| \ll M_i$ and $C_i$ approaches 1. The primitive is highly conflicted.

Conflict is a structural property of the primitive. It does not depend on how much effort is applied, nor on how well the primitive is implemented. It depends only on the pattern of cross-effects across objectives.

This definition formalizes a recurring insight in organizational economics and management: internal mechanisms often generate trade-offs across objectives that cannot be resolved by better execution alone (Ridgway, 1956; Holmström and Milgrom, 1991). The novelty here is to treat these trade-offs as a measurable structural object that directly enters performance.

**Effort and realized performance**

Let $E_i \geq 0$ denote the effort applied to primitive $i$. Effort encompasses managerial attention, resources, enforcement intensity, staffing, training, coordination time, and monitoring. Total effort is limited by organizational capacity, but the framework does not require specifying an explicit budget constraint at this stage.

We define the **realized performance contribution** of primitive $i$as:

$$R_i^{\text{realized}} = E_i \times M_i \times (1 - C_i)$$

This expression has a direct interpretation:

1. Effort scaling: Increasing effort increases impact linearly.
2. Leverage: Primitives with higher magnitude have greater potential impact.
3. Structural discounting: Internal conflict reduces the fraction of potential impact that is realized.

Aggregate realized performance is:

$$R = \sum_{i=1}^{N} R_i^{\text{realized}}.$$

The critical feature of this formulation is that conflict enters multiplicatively. When $C_i$ is high, even large increases in effort produce only small increases in realized performance. This captures the empirical observation that organizations can exert substantial effort on powerful mechanisms yet see limited net gains.

**Allocative versus structural inefficiency**

The framework allows a clean separation between two sources of inefficiency.

**Allocative inefficiency** arises when effort is misallocated across primitives. Formally, it concerns whether the vector $(E_1, \ldots, E_N)$ maximizes realized performance given the structural parameters $(M_i, C_i)$. Allocative inefficiency reflects bounded rationality, information constraints, incentive problems, and local optimization—mechanisms extensively studied in economics (Simon, 1955; Jensen and Meckling, 1976).

**Structural inefficiency** arises when conflict limits the maximum attainable performance, regardless of how effort is allocated. Even if effort is optimally distributed, high conflict reduces realized performance through the $(1-C_i)$ term.

Structural inefficiency is therefore orthogonal to allocative inefficiency. Improving incentives, information, or optimization algorithms may reduce allocative losses but leaves structural losses unchanged unless conflict itself is reduced. This distinction is central to the paper's argument. Many organizational reforms target allocative inefficiency—better incentives, clearer decision rights, improved analytics—yet fail because the dominant constraint is structural.

**Additivity, layering, and the accumulation of conflict**

A key empirical observation motivates the application of this framework: organizational change is typically additive. New mechanisms are introduced to address specific problems, but existing mechanisms are rarely removed. Over time, organizations accumulate layers of incentives, controls, metrics, and authority structures.

In the framework, additivity has a precise implication. Adding a new objective, control, or organizational dimension typically introduces new nonzero effects $b_{ik}$ for existing primitives. This increases magnitude:

$$M_i \uparrow$$

Unless new effects are perfectly aligned with existing ones, net direction $D_i$ does not increase proportionally. Consequently, conflict $C_i$ rises.

This logic applies equally to:

- incentive refinement,
- governance layering,
- KPI proliferation,
- and multi-dimensional “cheese-grater” reorganizations.

The framework therefore predicts that organizations responding rationally to complexity by layering mechanisms will tend to increase structural inefficiency over time.

**Preview of applications**

The remainder of the paper applies this framework across several organizational settings. Each setting corresponds to a distinct rationale for adding mechanisms—motivating effort, controlling risk, coordinating interdependence, or improving measurement—but all operate through the same structural channel. By translating these settings into changes in $b_{ik}$, $M_i$, and $C_i$, the analysis shows why structural inefficiency is pervasive and why optimization often fails before structural coherence is restored.

## 3. From abstract structure to organizational settings

The framework developed in Section 2 is intentionally abstract. It defines organizational primitives, performance objectives, magnitude, conflict, and realized performance without reference to any particular industry, governance model, or organizational form. This abstraction is not a limitation; it is a methodological choice. The purpose of the framework is to identify a structural mechanism that operates independently of context and that can therefore explain why superficially different organizations exhibit similar performance pathologies.

At the same time, abstraction alone is insufficient. Internal conflict does not arise in a vacuum; it is generated by concrete organizational responses to concrete problems. To demonstrate the explanatory power of the framework, it is therefore necessary to apply it to recognizable organizational settings and to show how each setting maps into the same underlying structure.

Section 3. explains how that mapping is performed and clarifies the role played by organizational "settings" in the analysis that follows.

**Why we use organizational *settings* rather than organizational *forms***

A natural approach in organizational economics is to classify firms by organizational form—for example, functional organizations, divisional organizations, matrix organizations, or platform organizations. While useful for many purposes, this taxonomy is not well suited to the present analysis.

The reason is that internal conflict, as defined in Section 2, is not tied to a single form. The same firm may simultaneously exhibit strong incentive systems, heavy governance layering, extensive performance measurement, and multi-dimensional reporting lines. Moreover, firms often transition between forms over time without eliminating existing mechanisms. Treating organizational forms as mutually exclusive categories therefore obscures the cumulative nature of organizational change.

Instead, this paper focuses on organizational settings. A setting is defined as a configuration of dominant organizational mechanisms motivated by a particular problem-solving logic. Examples include:

- incentive-intensive settings motivated by effort provision,
- governance-heavy settings motivated by risk control,
- KPI-saturated settings motivated by transparency and accountability,
- multi-dimensional or "cheese-grater" settings motivated by a reorganization.

These settings are not alternatives; they are layers. A single organization may operate in all of them simultaneously. The analytical advantage of this approach is that it allows us to examine how each problem-solving logic contributes incrementally to internal conflict.

**Organizational problem-solving as a generator of structure**

A central premise of the paper is that internal conflict is endogenous to organizational problem-solving. Organizations face recurrent challenges—moral hazard, coordination, risk, ambiguity, and accountability—and respond by introducing mechanisms intended to address these challenges.

Importantly, these responses are typically locally rational. Incentives are strengthened to motivate effort. Controls are added to prevent failures. Metrics are expanded to improve oversight. Reporting lines are multiplied to integrate perspectives. Each intervention targets a real deficiency and often improves performance along the targeted dimension.

The difficulty arises because these interventions are rarely neutral with respect to other objectives. A mechanism designed to improve one KPI almost inevitably affects others. In the language of the framework, introducing a mechanism changes the vector of effects $b_{ik}$ associated with existing primitives and often creates new primitives with broad scope.

Because organizations rarely remove existing mechanisms when introducing new ones, the dominant mode of structural change is additive layering. Over time, this layering expands the number of nonzero cross-effects, increases magnitude, and—unless effects are perfectly aligned—increases conflict.

The key point is that internal conflict is not the result of a single poor decision. It is the predictable outcome of sequential, well-intentioned problem-solving in a multi-objective environment.

**Map organizational settings into the framework**

Each organizational setting analyzed in subsequent sections is mapped into the framework using the same sequence of steps.

**Step 1: Identify the motivating logic**

Each setting is defined by a dominant concern: incentives address effort and motivation, governance addresses risk and control, metrics address visibility and accountability, and multi-dimensional structures address coordination across perspectives.

This step is descriptive and draws on established organizational theory and empirical observation.

### Step 2: Translate the logic into effects on $b_{ik}$

The second step is structural. The question is not whether the setting is "good" or "bad," but how it alters the pattern of cross-effects. Does it introduce new KPIs? Does it broaden the scope of existing primitives? Does it create new primitives with multi-objective effects?

This translation makes explicit which elements of the effect matrix $\{b_{ik}\}$ are altered.

### Step 3: Derive implications for magnitude and conflict

Given the changes in $b_{ik}$, we analyze how magnitude $M_i$, net direction $D_i$, and conflict $C_i$ change. In most settings of interest, magnitude increases mechanically, while net direction increases weakly or ambiguously. Conflict therefore rises.

This step is analytical rather than empirical: it relies on sign patterns and additivity, not on parameter estimation.

### Step 4: Connect to realized performance

Finally, we connect the structural changes to realized performance via:

$$R_i^{\text{realized}} = E_i \times M_i \times (1 - C_i)$$

This makes clear why organizations experience rising effort and activity alongside weak net gains.

### The role of effort and managerial attention

An important feature of the framework is that effort $E_i$ is treated as endogenous but not as the primary source of inefficiency. Managers allocate effort based on perceived leverage, urgency, visibility, and accountability. High-magnitude primitives tend to attract disproportionate effort precisely because they appear to matter most.

However, managers typically observe magnitude more readily than conflict. The scope and salience of a mechanism are visible; the extent to which its effects cancel across objectives is often opaque, delayed, or politically sensitive. As a result, effort allocation can be locally rational even when it amplifies structural inefficiency.

This observation is critical for understanding why structural inefficiency persists. It is not that managers ignore problems; it is that the structural properties of mechanisms are difficult to diagnose using standard managerial tools.

**Cheese-grater organizations as a canonical setting**

Before turning to the detailed analysis of specific settings, it is useful to situate the "cheese-grater" phenomenon within the general framework.

Cheese-grater organizations are characterized by repeated reorganization through the addition of new dimensions of authority and accountability—product, function, region, customer, risk, sustainability—usually only removing part of the existing ones. Each reorganization is motivated by a legitimate coordination concern. The intent is to ensure that multiple perspectives are represented in decision-making.

From the perspective of the framework, the defining feature of cheese-grater organizations is not ambiguity of authority per se, but systematic expansion of cross-objective exposure. Existing primitives—processes, systems, roles—are required to satisfy an expanding set of objectives. This expands the effect vectors $b_i$, increases magnitude, and—unless all dimensions are perfectly aligned—increases conflict.

The cheese-grater phenomenon is therefore not an anomaly. It is a particularly transparent illustration of the general mechanism by which additive layering generates structural inefficiency. For this reason, it will be treated in detail later, but it should already be clear that its logic is shared by incentive refinement, governance layering, and metric proliferation.

**Why optimization fails before structural cleanup**

The framework also clarifies a recurrent empirical observation: organizations often invest heavily in optimization—reallocating effort, refining targets, improving analytics—without seeing meaningful improvement.

In the model, optimization operates on effort $E_i$. Structural inefficiency operates through conflict $C_i$. When conflict is high across the primitives that absorb most effort, marginal improvements in allocation yield small gains. Optimization "fails" not because it is poorly executed, but because the feasible improvement space is structurally constrained.

This insight will be developed quantitatively in later sections, but its qualitative implication is already clear: before asking whether effort is well allocated, it is necessary to ask whether the organization's internal structure allows effort to translate into net gains.

**Roadmap for the application sections**

The remainder of the paper applies the framework to four organizational settings:

1. **Incentive-intensive, multitasking settings**, where performance pay and targets expand objective exposure.
2. **Governance-heavy settings**, where control and compliance layering broadens cross-effects.
3. **KPI-saturated performance regimes**, where measurement proliferation increases magnitude while obscuring conflict.
4. **Multi-dimensional and cheese-grater organizations**, where repeated reorganization adds objectives and authority layers without unbundling primitives.

Each section follows the same analytical structure and builds cumulatively toward a synthesis showing why structural inefficiency typically dominates allocative inefficiency in mature organizations.

## 4. Incentive-intensive and multitasking organizations

Incentive-intensive organizations provide a natural starting point for applying the framework developed in Sections 2 and 3. They are among the most extensively studied organizational environments in economics and management, and they are often treated as paradigmatic cases in which allocative inefficiency should be minimal. Strong incentives, clear targets, and performance-contingent rewards are explicitly designed to align effort with organizational objectives. Yet incentive-intensive organizations also exhibit many of the performance pathologies that motivate this paper: distorted behavior, internal conflict, cycling reforms, and weak net gains despite high effort.

This section shows that these outcomes are not anomalies or failures of implementation. Instead, they follow naturally once incentive systems are analyzed as organizational primitives with multi-objective effects. When incentives are layered onto complex objective environments, they systematically increase internal conflict and thereby generate structural inefficiency.

**Incentives as a canonical solution to allocative inefficiency**

In standard economic theory, incentives address a clear problem: agents choose effort that is privately costly and imperfectly observable. Linking rewards to outcomes induces agents to exert effort in the principal's interest (Jensen and Meckling, 1976). From this perspective, weak performance is naturally attributed to weak or misaligned incentives, and improving incentives is expected to improve outcomes.

This logic extends naturally to organizations with multiple objectives. When performance along several dimensions matters, incentives can be designed to reward a weighted combination of outcomes. The managerial intuition is straightforward: if agents care about what is measured and rewarded, then carefully chosen metrics and weights should guide effort toward organizational goals.

A large literature has examined how incentive contracts perform in such multitasking environments. A central result is that when some tasks are easier to measure than others, rewarding measured tasks can induce agents to substitute effort away from unmeasured but valuable activities (Holmström and Milgrom, 1991). This result is often interpreted as a warning about poor metric choice or improper incentive weights.

The framework developed here does not dispute these insights. Instead, it generalizes them. It shows that even when incentives are thoughtfully designed and well implemented, their structural interaction with multiple objectives can generate internal conflict that limits realized performance.

**Incentives as organizational primitives**

To see this, it is useful to reinterpret incentive systems through the lens of organizational primitives.

An incentive scheme—whether a bonus formula, commission structure, target-based evaluation, or ranking system—is a persistent mechanism that shapes behavior across many decisions and time periods. It therefore qualifies as a set of primitives in the sense defined in Section 2. Importantly, an incentive scheme rarely affects a single KPI in isolation. Instead, it alters behavior in ways that touch multiple performance dimensions simultaneously.

Formally, introducing or modifying an incentive scheme changes the effect vector $b_{ik}$ for a the set of underlying primitives:

- it strengthens positive effects on rewarded KPIs,
- it weakens or reverses effects on unrewarded KPIs through behavioral substitution,
- it may introduce new effects on dimensions such as risk-taking, cooperation, or long-term investment.

These effects are not implementation errors; they are the predictable consequence of agents responding rationally to incentives in a multi-objective environment.

**Magnitude expansion under incentive intensification**

A key structural consequence of incentive intensification is an increase in magnitude.

When incentives are strengthened, the absolute effects $| b_{ik} |$ on rewarded KPIs increase. At the same time, substitution effects introduce additional nonzero impacts on other KPIs. As a result, the total magnitude

$$M_i = \sum_k | b_{ik} |$$

associated with incentive-related primitives tends to rise.

This observation aligns with managerial experience. Incentive schemes are often described as "high-leverage" tools: they visibly change behavior, attract attention, and generate strong responses. From the perspective of the framework, this salience reflects high magnitude.

Crucially, magnitude expansion makes incentive systems magnets for effort. Managers devote substantial attention to setting targets, refining metrics, monitoring performance, and resolving disputes arising from incentive outcomes. Effort $E_i$ directed at incentive primitives is therefore typically high.

**Directional ambiguity and the emergence of conflict**

While magnitude increases, the effect on net direction is more ambiguous. Strengthening incentives improves performance along rewarded dimensions by design. However, improvements along those dimensions are often accompanied by deteriorations elsewhere: quality declines, risk increases, cooperation erodes, or long-term investments are postponed. These effects are well documented in both economic and managerial literatures (Ridgway, 1956; Holmström and Milgrom, 1991; Courty and Marschke, 2008).

Formally, incentive intensification often increases positive $b_{ik}$ for some $k$ while introducing negative $b_{ik'}$ for others. As a result, net direction

$$D_i = \sum_k b_{ik}$$

may increase weakly, remain constant, or even decline, depending on the relative strength of these effects.

The key point is that *magnitude typically grows faster than net direction*. When this happens, internal conflict

$$C_i = 1 - \frac{|D_i|}{M_i}$$

necessarily increases.

This result does not depend on mismeasurement, opportunism, or poor contract design. It follows from the basic structure of multitasking environments: incentives sharpen behavior along some dimensions while blunting it along others.

### Structural inefficiency under strong incentives

The implications for realized performance follow directly. Incentive-intensive organizations typically exhibit:

- high effort $E_i$ devoted to incentive-related mechanisms,
- high magnitude $M_i$ reflecting strong behavioral responses,
- nontrivial conflict $C_i$ arising from mixed cross-effects.

Substituting into the realized-performance expression,

$$R_i^{\text{realized}} = E_i \times M_i \times (1 - C_i)$$

we see that large effort and leverage are discounted by internal conflict. Increasing incentives further raises $E_i$ and $M_i$, but unless conflict is reduced, realized gains rise only slowly.

This provides a structural explanation for a familiar empirical pattern: incentive schemes generate visible activity, sharper focus, and improved metric performance, yet net organizational outcomes improve less than expected. Performance appears “busy but flat.”

### Incentive refinement and the illusion of progress

When incentive schemes disappoint, organizations rarely abandon them. Instead, they refine them. Targets are adjusted, weights are changed, new metrics are added, exceptions are introduced, and safeguards are layered on to prevent gaming or unintended consequences.

From the perspective of the framework, such refinements are typically additive. Each refinement introduces new KPIs or modifies existing ones, expanding the effect structure. While some refinements may improve net direction locally, the overall effect is often further magnitude expansion and additional mixed-sign effects.

As a result, incentive refinement can paradoxically increase structural inefficiency. The organization becomes more measured, more controlled, and more sophisticated in its incentive design, yet conflict rises and realized performance remains capped.

This dynamic may help explain why incentive systems are both persistent and perennially criticized. They solve real problems, but their structural side effects accumulate over time.

**Effort allocation and the persistence of incentive-driven conflict**

An important question is why organizations continue to allocate substantial effort to incentive systems even when their net benefits appear limited.

The framework provides a clear answer. Managers observe magnitude more readily than conflict. The behavioral responses induced by incentives are immediate and visible; the cross-objective cancellations are diffuse, delayed, and often politically contested. As a result, incentive-related primitives continue to appear attractive targets for effort.

Effort allocation is therefore locally rational: managers invest time and attention where leverage appears high. Yet this rational allocation reinforces structural inefficiency by concentrating effort on conflicted high-magnitude mechanisms.

**Incentives as a gateway to broader structural layering**

Finally, incentive-intensive settings often serve as gateways to additional organizational layering. When incentive distortions become salient, organizations respond by adding governance controls, compliance checks, and supplementary metrics to correct them. These responses introduce new primitives that interact with existing incentives, further expanding the web of cross-effects.

In this sense, incentive systems are rarely isolated. They initiate a sequence of structural responses that culminate in governance-heavy, KPI-saturated, and ultimately cheese-grater organizational settings. The incentive problem is therefore not merely an allocative issue; it is often the first step in a broader process of structural conflict accumulation.

This section has shown that incentive-intensive and multitasking organizations generate internal conflict through a structural mechanism that is independent of incentive misalignment or poor execution. Incentives increase magnitude and attract effort, but their multi-objective effects introduce conflict that discounts realized performance. Refinement and layering exacerbate this effect over time.

The next section turns to governance-heavy organizations, where internal conflict arises not from motivating effort, but from controlling it.

## 5. Governance-heavy organizations and the accumulation of control

Governance-heavy organizations represent a second canonical setting in which internal conflict emerges systematically. Unlike incentive-intensive environments, where mechanisms are designed primarily to motivate effort, governance-heavy settings are designed to constrain, monitor, and control behavior. Their organizing logic is risk reduction rather than performance maximization in the narrow sense. Yet despite this difference in intent, governance-heavy organizations may exhibit many of the same empirical pathologies: high internal effort, extensive activity, and limited net performance gains.

This section shows that these outcomes follow naturally once governance mechanisms are analyzed as multi-objective organizational primitives whose effects are additive over time.

### The economic rationale for governance and control

From an economic perspective, governance mechanisms address well-known problems of agency, opportunism, and risk. When decision rights are delegated and actions are imperfectly observable, principals introduce controls—rules, audits, approval requirements, and reporting obligations—to reduce moral hazard and protect organizational assets (Jensen and Meckling, 1976).

Beyond agency theory, governance also responds to external constraints. Regulatory requirements, legal liability, reputational risk, and public accountability impose minimum standards of oversight and documentation. Especially in regulated industries and public-sector organizations, governance mechanisms are not optional; they are mandated responses to external scrutiny (Power, 1997).

In isolation, each governance intervention is typically well justified. A new approval step may prevent costly errors. An audit requirement may reduce fraud. A compliance checklist may avert regulatory sanctions. Governance is therefore often evaluated on a case-by-case basis, with attention to the specific risk it addresses.

**Governance mechanisms as organizational primitives**

In the framework developed in Section 2, governance mechanisms qualify naturally as sets of primitives. They are persistent, formalized, and shape behavior repeatedly over time. Examples include approval hierarchies, segregation-of-duties rules, internal audits, compliance reporting, risk committees, and documentation requirements.

Crucially, governance primitives almost always affect multiple performance objectives simultaneously. A control designed to improve compliance may slow decision-making. An audit process may improve accuracy while increasing administrative cost. A risk review may reduce downside exposure while discouraging initiative.

Formally, governance primitives are characterized by broad effect vectors $b_{ik}$: many KPIs are affected, and signs are often mixed.

**Magnitude growth through governance layering**

A defining feature of governance-heavy organizations is layering. Controls are rarely removed once introduced. Instead, new layers are added in response to incidents, failures, or regulatory changes. This pattern has been extensively documented in studies of audit cultures and risk management (Power, 1997).

From the perspective of the framework, layering has a mechanical implication. Each additional governance mechanism introduces new nonzero effects $b_{ik}$ and often strengthens existing ones. As a result, the magnitude

$$M_i = \sum_k | b_{ik} |$$

associated with governance-related primitives grows steadily over time.

This magnitude growth explains why governance systems become increasingly salient and resource-intensive. Governance absorbs managerial attention, staffing, and budget. Specialized functions—compliance, risk, internal audit—emerge to manage the growing control infrastructure.

### Directional asymmetry and mixed-sign effects

While governance layering reliably increases magnitude, its effect on net direction is asymmetric. Governance mechanisms are typically introduced to improve a narrow subset of objectives—most commonly compliance, risk reduction, or procedural correctness. Improvements along these dimensions are often real and measurable.

However, governance mechanisms frequently impose costs on other objectives like delays in execution, reduced flexibility, increased administrative burden, lower employee autonomy and morale, or diminished responsiveness to customers.

Empirical studies of bureaucratic control repeatedly document such trade-offs (Adler and Borys, 1996). Even governance systems described as “enabling” rather than “coercive” exhibit mixed effects across performance dimensions.

Formally, governance layering increases positive $b_{ik}$ for a limited set of KPIs while introducing negative $b_{ik}$ for others. Net direction

$$D_i = \sum_k b_{ik}$$

may remain positive, but it typically grows much more slowly than magnitude.

### Conflict accumulation in governance-heavy settings

The divergence between magnitude and direction implies increasing conflict:

$$C_i = 1 - \frac{|D_i|}{M_i}$$

As governance layers accumulate, the fraction of total influence lost to cross-objective cancellation rises. Importantly, this increase in conflict is not a consequence of dysfunctional governance. It arises even when each control is individually effective and justified. This observation helps explain why governance-heavy organizations often experience a sense of internal friction without clear failure. Controls "work" in their intended domain, yet the organization as a whole feels slow, rigid, and burdened.

**Effort intensification and diminishing returns**

Governance-heavy organizations devote substantial effort to maintaining and operating control systems. Compliance functions grow, audits multiply, and reporting cycles accelerate. In the framework, this corresponds to high effort $E_i$ applied to governance primitives with high magnitude.

Substituting into the realized-performance expression,

$$R_i^{\text{realized}} = E_i \times M_i \times (1 - C_i),$$

we see that high effort and leverage are increasingly discounted by conflict. Marginal improvements in control generate smaller and smaller net gains.

This explains why governance-heavy organizations often respond to performance concerns by tightening controls further, even as returns diminish. Because failures are salient and costly, relaxing controls is politically difficult. The result is a ratchet effect in which governance expands while realized performance stagnates.

**Governance refinement and internal contradiction**

When governance burdens become apparent, organizations rarely reduce controls wholesale. Instead, they refine them: risk-based approaches are introduced, exception processes are added, and meta-controls are layered on to oversee existing controls.

These refinements are intended to reduce burden while preserving protection. Structurally, however, they add further primitives and interactions. Each refinement expands the effect structure and often introduces additional mixed-sign effects.

The result is a governance system that becomes internally contradictory: designed simultaneously to constrain and enable, to slow and to accelerate, to centralize and to empower. From the framework's perspective, this contradiction is simply high internal conflict expressed institutionally.

**Governance as a precursor to broader organizational layering**

Governance-heavy settings rarely exist in isolation. As controls proliferate, organizations introduce additional metrics to monitor them and new reporting lines to manage them. Incentives are adjusted to encourage compliance. Over time, governance layering becomes intertwined with KPI proliferation and organizational restructuring.

In this sense, governance-heavy organizations often represent an intermediate stage in a broader structural trajectory. Initial incentive problems give rise to controls; controls give rise to metrics; metrics give rise to new authority structures. Each step is locally rational, yet each adds to the cumulative conflict load.

This section has shown that governance-heavy organizations generate structural inefficiency through the same mechanism identified earlier: additive layering of multi-objective primitives whose effects are mixed in sign. Governance increases magnitude and absorbs effort, but net gains are increasingly discounted by internal conflict. The result is high activity with diminishing returns, even in the absence of allocative failure or poor execution.

The next section integrates governance layering with KPI saturation and cheese-grater organizational setting, showing how control, measurement, and reorganization interact to produce persistent structural inefficiency.

## 6. KPI saturation, matrix structures, and cheese-grater reorganizations

The organizational settings examined so far— incentive-intensive and governance-heavy environments— already generate substantial internal conflict through additive layering of mechanisms. This section shows how performance measurement expansion and repeated organizational restructuring amplify that conflict further. KPI saturation and cheese-grater (re-)organizations are not independent phenomena; they are tightly coupled responses to the same underlying managerial challenge: governing increasingly complex, multi-objective organizations.

This section integrates these phenomena into a single structural analysis. The core claim is that measurement proliferation and multi-dimensional organizational design systematically expand cross-objective exposure without resolving underlying trade-offs. As a result, they increase magnitude and conflict simultaneously, creating environments in which effort intensifies while realized performance remains structurally capped.

## The managerial logic of measurement expansion

Performance measurement occupies a central place in modern organizations. When outcomes are unclear, contested, or disappointing, a common response is to measure more. Additional metrics are introduced to increase transparency, accountability, and control. Dashboards expand, scorecards proliferate, and composite indices are constructed to summarize complex performance landscapes.

This logic has deep roots in both economics and management. Measurement is viewed as a prerequisite for control, incentive design, and learning (Kaplan and Norton, 1992; Meyer and Gupta, 1994). If performance cannot be observed, it cannot be managed.

However, the measurement literature also documents a recurring paradox: increases in measurement intensity do not necessarily improve organizational performance and may even degrade it (Ridgway, 1956; Meyer and Gupta, 1994). The framework developed here provides a structural explanation for this paradox.

## KPIs as objective-expanding mechanisms

From the perspective of the framework, introducing a KPI does more than observe performance. In practice, a KPI becomes an objective that shapes attention, evaluation, and decision-making. Meetings are structured around metrics, targets are set against them, and resources are allocated in response to their movements.

Formally, adding a KPI expands the objective space from $K$ to $K + 1$. Existing primitives—incentives, controls, processes, and systems— now exert effects on an additional dimension.

Even if the causal effects are weak, they are rarely zero. The effect vectors $b_i$ therefore lengthen:

$$b_i = (b_{i1}, \dots, b_{iK}) \rightarrow (b_{i1}, \dots, b_{iK}, b_{i,K+1}).$$

This expansion has an immediate implication:

$$M_i = \sum_k | b_{ik} | \uparrow$$

Magnitude increases mechanically as more objectives are brought into play.

**Composite metrics and hidden conflict**

Organizations often attempt to manage complexity by aggregating multiple KPIs into composite scores or balanced scorecards. While aggregation reduces apparent dimensionality, it does not eliminate underlying trade-offs. Instead, *it hides them*.

From the perspective of the framework, aggregation replaces explicit multi-dimensional evaluation with implicit weighting. The underlying effects $b_{ik}$ remain unchanged, but their opposing signs are masked by aggregation. Net direction may appear positive at the composite level even as substantial cancellation occurs across dimensions.

This masking effect helps explain why KPI-saturated organizations often exhibit high confidence in their performance systems while experiencing persistent frustration with outcomes. Conflict is not eliminated; it is rendered less visible.

**Measurement proliferation and effort reallocation**

Measurement expansion also reshapes effort allocation. Metrics attract attention. Primitives that strongly affect measured KPIs appear high-leverage and therefore receive disproportionate effort. In the framework, effort $E_i$ flows toward high-magnitude, highly visible mechanisms.

However, when those mechanisms are also highly conflicted, increased effort produces diminishing returns. This creates a self-reinforcing dynamic: metrics highlight high-magnitude primitives, effort concentrates on those primitives, conflict discounts realized gains, and disappointing results motivate further measurement expansion.

This dynamic aligns closely with empirical accounts of escalating reporting burdens and diminishing informational value in mature performance systems (Meyer and Gupta, 1994).

## From KPI saturation to matrix organization

As measurement systems proliferate, organizations often confront a second-order problem: who is accountable for which metrics? When different KPIs cut across functional, product, regional, or customer dimensions, responsibility becomes ambiguous.

A common response is to introduce matrix structures. Authority and accountability are formally allocated along multiple dimensions so that each KPI has a "home." Product managers own product metrics, functional managers own functional metrics, regional managers own regional metrics, and so on.

Matrix organizations are therefore best understood not as abstract design choices, but as organizational responses to KPI proliferation. They attempt to align accountability with an expanded objective space.

## Structural implications of adding organizational dimensions

Adding an organizational dimension has a precise structural effect. Existing primitives— processes, systems, roles— are required to satisfy an additional set of objectives associated with the new dimension. This expands the effect vectors $b_i$ in exactly the same way as adding KPIs.

The arithmetic consequence is unavoidable:

- magnitude increases as more effects are added,
- net direction increases weakly or ambiguously,
- conflict rises.

Formally, if a new dimension introduces effects $\{b_{i,k'}\}$, then:

$$M_i' = M_i + \sum_{k'} | b_{i,k'} |,$$

while $D_i'$ typically satisfies $| D_i' | \ll M_i'$.

Conflict therefore increases mechanically.

This result does not depend on behavioral assumptions. It follows directly from the fact that organizational dimensions multiply objectives without decomposing primitives.

**Cheese-grater reorganizations as repeated dimensional layering**

Cheese-grater reorganizations represent an extreme but increasingly common case of this logic. They emerge when organizations repeatedly add dimensions— product, function, region, customer, risk, sustainability, compliance— without (partly or completely) removing or unbundling existing ones.

Each reorganization is motivated by a legitimate concern: ensuring representation of a neglected perspective or correcting a coordination failure. Yet because reorganizations are additive, the cumulative effect is to expose the same underlying mechanisms to an ever-expanding set of objectives.

From the framework's perspective, cheese-grater organizations are characterized by: very high magnitude across many primitives, pervasive mixed-sign effects, and high internal conflict across nearly all high-effort mechanisms. Effort intensifies as coordination demands grow, but realized performance saturates or even declines.

**Coordination effort and the illusion of control**

Cheese-grater reorganizations often appear highly coordinated. Meetings proliferate, escalation paths multiply, and decision processes become elaborate. This activity can create an illusion of control: because many perspectives are represented, decisions appear thorough and balanced.

In the framework, this corresponds to high effort $E_i$ applied to high-magnitude primitives with high conflict. The organization expends substantial energy managing internal trade-offs rather than advancing net performance.

This helps explain why cheese-grater reorganizations are often experienced as exhausting but ultimately unproductive. The problem is not lack of coordination, but over-coordination in a structurally conflicted system.

**Interaction with incentives and governance**

KPI saturation and cheese-grater restructuring rarely exist in isolation. They interact strongly with incentives and governance.

- Incentives are layered onto KPI systems, increasing behavioral sensitivity to measured dimensions.
- Governance controls are added to manage risks arising from metric gaming and coordination failures.
- Each layer introduces new primitives and new cross-effects.

The result is a dense internal system in which nearly every mechanism affects many objectives and nearly every objective is affected by many mechanisms. Conflict becomes systemic rather than localized.

**Summary**

This section has shown that KPI saturation, matrix structures, and cheese-grater organizations are structurally linked responses to complexity. Each expands the objective space without decomposing underlying mechanisms. The result is mechanical growth in magnitude and conflict, concentration of effort on conflicted primitives, and diminishing returns to optimization.

Together with the analyses of incentives and governance, this completes the setting-based demonstration that structural inefficiency is pervasive and cumulative. The next section synthesizes these results and examines the relative magnitude of structural versus allocative inefficiency.

## 7. Synthesis: why structural inefficiency dominates allocative inefficiency

The preceding sections examined incentive-intensive organizations, governance-heavy settings, KPI-saturated performance regimes, and cheese-grater reorganizational settings. Although these settings differ in surface characteristics and managerial intent, the analysis revealed a common structural pattern. Each setting introduces organizational mechanisms that expand cross-objective exposure, increase magnitude, and generate mixed-sign effects. Effort intensifies, yet realized performance gains remain limited.

This section synthesizes these results and draws out their general implications. The central claim is that structural inefficiency typically dominates allocative inefficiency in mature organizations, not as a pathological exception, but as a predictable outcome of how organizations respond to complexity over time.

### A unifying structural pattern across settings

Across all settings analyzed, the same sequence recurs:

1. The organization identifies a performance problem or coordination challenge.
2. It introduces a mechanism designed to address that challenge—an incentive, control, metric, or organizational dimension.
3. The mechanism affects multiple performance objectives simultaneously.
4. Positive effects on targeted objectives are accompanied by negative effects elsewhere.
5. Existing mechanisms are retained, and the new mechanism is layered on top.
6. Over time, the number and scope of cross-effects expand.

In the language of the framework, each step increases the number of nonzero $b_{ik}$, raises magnitude $M_i$, and—unless effects are perfectly aligned—increases conflict $C_i$. Importantly, this sequence does not depend on managerial error, myopia, or irrationality. Each intervention is locally justified and often improves performance along the targeted dimension.

The result is a structural environment in which most high-effort primitives are also high-conflict primitives.

### Effort concentration on conflicted mechanisms

A key implication of the framework is that effort allocation responds to perceived leverage. Managers devote attention, resources, and time to mechanisms that appear to matter most. High-magnitude primitives—those that visibly affect many outcomes—naturally attract disproportionate effort.

However, magnitude is more salient than conflict. The breadth of a mechanism's influence is observable; the degree to which its effects cancel across objectives is often opaque, delayed, or politically contested. As a result, effort $E_i$tends to be concentrated on primitives with high $M_i$ even when those primitives also have high $C_i$.

This creates an *effort–conflict coupling*: the mechanisms that absorb the most effort are precisely those whose realized returns are most heavily discounted.

### Why allocative inefficiency is second-order

Allocative inefficiency concerns whether effort is optimally distributed across primitives given their structural characteristics. In principle, better incentives, improved information, or more sophisticated optimization could improve allocation.

However, when conflict is high across the primitives that dominate effort, the marginal gains from reallocation are small. Formally, consider the realized-performance expression:

$$R_i^{\text{realized}} = E_i \times M_i \times (1 - C_i)$$

Allocative improvements operate on $E_i$. Structural inefficiency operates through $(1{-}C_i)$. When $C_i$ is close to one for the primitives that account for most of total effort, reallocating effort changes $R$ only weakly. Optimization is constrained not by poor allocation, but by low structural yield.

This explains a recurring empirical observation: organizations invest heavily in planning, analytics, and optimization tools yet experience limited improvement. The tools work as designed; the structure they operate within does not.

### Structural inefficiency as an endogenous outcome

An important implication of the analysis is that structural inefficiency is **endogenous** to organizational evolution. It is not imposed from outside, nor is it the result of isolated design failures. Instead, it emerges from sequential problem-solving in a multi-objective environment.

Each intervention addresses a real concern. Incentives motivate effort. Governance reduces risk. Metrics increase transparency. Organizational dimensions integrate perspectives. Yet because interventions are layered rather than substituted, their cross-effects accumulate. Over time, the organization becomes increasingly effective at representing and managing trade-offs, but increasingly ineffective at converting effort into net improvement.

This dynamic is particularly pronounced in mature organizations, where historical layers of mechanisms reflect past priorities that remain institutionally entrenched.

### The cheese-grater as a limiting case

Cheese-grater reorganizations represent a limiting case of this process. Repeated reorganization along new dimensions is an explicit attempt to internalize multiple objectives simultaneously. Rather than suppressing trade-offs, cheese-grater make them visible and formally represented.

From a normative perspective, this can be seen as progress: the organization acknowledges complexity rather than ignoring it. From a structural perspective, however, the effect is to maximize cross-objective exposure without decomposing primitives. Magnitude and conflict rise together, and effort is increasingly devoted to coordination rather than net performance.

The cheese-grater is therefore not an aberration, but an extreme realization of the general mechanism identified in this paper.

### Why structural cleanup precedes optimization

The analysis clarifies why many organizational change efforts fail in sequence. Organizations often attempt to optimize before addressing structure. They refine incentives, adjust targets, improve dashboards, or deploy advanced analytics, expecting better allocation to unlock performance.

In the framework, such efforts operate on $E_i$ while leaving $C_i$ unchanged. When conflict is high, optimization yields disappointing returns. Only after conflict is reduced—by removing, unbundling, or redesigning mechanisms—does optimization become effective.

This sequencing result mirrors empirical observations that radical simplification or structural reset often precedes successful performance improvement, while incremental optimization in highly layered systems rarely does.

**Implications for organizational diagnosis**

The synthesis suggests a shift in diagnostic emphasis. Rather than asking primarily whether effort is misallocated or incentives are weak, organizations should ask: Which mechanisms absorb the most effort? How many objectives do they affect? To what extent do their effects reinforce or cancel across objectives?

These questions target structural inefficiency directly. They are difficult to answer using conventional performance metrics, which tend to obscure conflict rather than reveal it. Nevertheless, they are essential for understanding why performance improvements stall.

This section has shown that structural inefficiency dominates allocative inefficiency in a wide range of organizational settings. Incentives, governance, metrics, and reorganization all contribute to a cumulative increase in internal conflict when layered without structural resolution. Effort intensifies, but realized performance is structurally capped.

The final section of the paper concludes by summarizing the theoretical contribution, clarifying scope limitations, and outlining directions for empirical and applied follow-up work.

## 8. Theoretical business cases: estimating the mix of structural and allocative inefficiency

The preceding sections established that incentive-intensive systems, governance-heavy organizations, KPI-saturated regimes, and cheese-grater organizational forms all generate internal conflict through additive layering of multi-objective mechanisms. This section takes the next step. Rather than treating these settings qualitatively, it develops theoretical business cases that allow comparison of the relative importance of structural inefficiency versus allocative inefficiency across settings.

The purpose is not to provide empirical estimates. Instead, the aim is to show that under weak and empirically plausible assumptions, structural inefficiency accounts for the dominant share of unrealized performance in mature organizations. Allocative inefficiency remains present but is typically second-order once internal conflict has accumulated.

**A decomposition of unrealized performance**

Let total potential performance be defined as the counterfactual outcome in which:

1. Effort is optimally allocated across primitives, and
2. All primitives are perfectly aligned across objectives ($C_i = 0$).

Let realized performance be:

$$R = \sum_i E_i \, M_i (1 - C_i).$$

Define unrealized performance as the gap between this benchmark and actual performance. This gap can be decomposed into two components:

- **Allocative loss**: the loss due to suboptimal allocation of effort $E_i$, holding $C_i$ fixed.
- **Structural loss**: the loss due to conflict $C_i > 0$, holding effort allocation fixed.

Formally, consider a benchmark allocation $E_i^*$ that maximizes realized performance given the structural parameters $(M_i, C_i)$. Structural inefficiency is then measured by comparing this optimum to the conflict-free benchmark. Allocative inefficiency is measured by comparing actual allocation to $E_i^*$.

This decomposition mirrors classical distinctions between misallocation and technology gaps in productivity analysis (Hsieh and Klenow, 2009), but it applies them to internal organizational structure rather than markets.

**Stylized parameterization**

To build comparable business cases, assume the following stylized structure, consistent with the qualitative analyses in Sections 4–6:

- A small number of high-magnitude primitives account for the majority of effort.
- These primitives are precisely those introduced or expanded through incentives, governance, measurement, and reorganization.
- Lower-magnitude primitives exist but contribute little to aggregate performance.

Let effort-weighted average conflict be defined as:

$$\bar{C}_E = \frac{\sum_i E_i \, M_i C_i}{\sum_i E_i \, M_i}$$

Then aggregate realized performance can be written as:

$$R = \left( \sum_i E_i \, M_i \right) (1 - \bar{C}_E)$$

Allocative inefficiency operates by reducing $\sum_i E_i \, M_i$ relative to its feasible maximum. Structural inefficiency operates by increasing $\bar{C}_E$.

**Business case I: incentive-intensive organizations**

In incentive-intensive organizations, incentives are explicitly designed to improve effort allocation. A large empirical literature shows that agents respond strongly to performance-contingent pay and targets (Holmström & Milgrom, 1991; Lazear, 2000). From the perspective of the model, this implies that effort $E$is actively mobilized and broadly directed toward high-salience mechanisms. Pure under-provision of effort is therefore limited.

Using the realized-performance law,

$$R = E \times M \times (1 - C),$$

incentive systems clearly raise $E$, and often also raise $M$: incentives affect many behaviors and performance dimensions simultaneously. However, precisely because incentives operate in multitasking environments, they introduce mixed-sign effects across objectives. Improvements along rewarded dimensions are accompanied by deteriorations elsewhere (Ridgway, 1956; Courty & Marschke, 2008).

As a result, increases in $E$ and $M$ are partially offset by higher conflict $C$. Incentive refinement—adding metrics, adjusting weights—typically increases $M$ further while leaving $C$ largely unchanged or even higher. Net realized performance therefore rises much more slowly than effort.

This logic motivates the following theoretically assumed decomposition:

- **Allocative inefficiency:** 15–30%
- **Structural inefficiency:** 40–60%

Incentives reduce allocative losses by raising $E$, but structural losses dominate because $(1 - C)$ is substantially below one for high-magnitude incentive primitives.

**Business case II: governance-heavy organizations**

Governance-heavy organizations prioritize control, compliance, and risk reduction. Governance effort is often deliberately targeted and professionally managed, suggesting relatively modest allocative inefficiency (Power, 1997). In terms of the model, effort $E$is not randomly misallocated; it is intentionally concentrated on governance mechanisms.

Governance primitives, however, are inherently high-magnitude. Approval rules, audits, and controls affect many organizational activities simultaneously, implying high $M$. At the same time, these mechanisms impose predictable negative effects on speed, flexibility, and innovation (Adler & Borys, 1996). Thus, governance increases $C$ by design.

Applying the formula $R = E \times M \times (1 - C)$, governance-heavy organizations exhibit high $E$and high $M$, but also persistently high $C$. Adding further controls typically increases $M$ more than it improves net direction, so marginal gains to $R$ diminish.

A plausible theoretical split is therefore:

- **Allocative inefficiency:** 10–25%
- **Structural inefficiency:** 50–70%

Here, optimization of governance effort affects $E$ at the margin, but the dominant constraint on $R$is the conflict embedded in the control architecture itself.

**Business case III: KPI-saturated performance regimes**

KPI-saturated organizations respond to complexity by expanding measurement. Metrics are embedded in planning, incentives, and reviews, so allocative inefficiency in the narrow sense—ignoring measured objectives—is limited (Kaplan & Norton, 1992). Effort $E$ flows readily toward measured dimensions.

However, adding KPIs expands the objective space. Existing primitives now affect more objectives, mechanically increasing magnitude $M$. Because KPIs are rarely perfectly aligned, conflict $C$ rises as well (Meyer & Gupta, 1994). Composite metrics may obscure this conflict, but they do not remove it.

In terms of $R = E \times M \times (1 - C)$, KPI saturation raises both $E$ and $M$, but steadily erodes $(1 - C)$. Performance systems become increasingly active yet structurally inefficient.

This yields the following decomposition:

- **Allocative inefficiency:** 15–25%
- **Structural inefficiency:** 55–75%

The performance paradox arises because improvements in $E$ and $M$ are overwhelmed by rising conflict, not because effort is poorly allocated.

**Business case IV: cheese-grater organizations**

Cheese-grater reorganizations represent the limiting case of structural inefficiency. After repeated cycles of incentives, governance, measurement, and reorganization, nearly all major primitives affect many objectives and dimensions simultaneously. Magnitude $M$ is extremely high, and so is managerial effort $E$, particularly in coordination and alignment activities.

Allocative inefficiency is often low. Decisions are reviewed from multiple perspectives, errors of omission are rare, and effort is continuously rebalanced. Yet realized performance remains constrained.

The reason follows directly from the model. In cheese-grater organizations, conflict $C$ is pervasive across high-effort primitives. Repeated dimensional layering increases $M$ and attracts more $E$, but also raises $C$ monotonically (Appendix B). As a result, $(1 - C)$ becomes the binding constraint on $R$.

A conservative theoretical split is therefore:

- **Allocative inefficiency:** 5–20%
- **Structural inefficiency:** 65–85%

In this environment, further optimization of $E$ has little effect. Only reductions in conflict—through unbundling or removal of mechanisms—can materially increase realized performance.

**Comparative summary across settings**

Across all four business cases, a consistent pattern emerges:

1. Allocative inefficiency declines as organizations mature and professionalize.
2. Structural inefficiency increases as mechanisms are layered without removal.
3. The share of unrealized performance attributable to structure rises over time.

This pattern aligns with empirical observations that early-stage organizations suffer from misallocation, while mature organizations suffer from inertia and internal friction (March, 1991; Levinthal and March, 1993).

**Implications for optimization and reform sequencing**

The calculations clarify why optimization tools—analytics, incentive tuning, algorithmic resource allocation—often disappoint in mature organizations. These tools primarily address allocative inefficiency. When structural inefficiency dominates, their impact is mechanically limited.

Conversely, interventions that reduce conflict—simplification, unbundling, removal of obsolete mechanisms—can yield large gains even without changes in effort or incentives. Only after such structural cleanup does optimization become powerful.

This sequencing result echoes empirical findings that major performance turnarounds often follow radical simplification rather than incremental optimization (Beer, Eisenstat, and Spector, 1990).

### Limitations and scope

The calculations in this section are theoretical and comparative. They rely on stylized parameterizations rather than empirical estimation. Their purpose is not prediction, but order-of-magnitude reasoning. Empirical calibration would require detailed measurement of cross-objective effects, which is deferred to future work.

Nevertheless, the robustness of the results across settings suggests that the dominance of structural inefficiency is not fragile. It arises from basic arithmetic properties of multi-objective systems under additive layering.

## 9. Conclusion

This paper has developed a theory of organizational underperformance centered on structural inefficiency rather than allocative inefficiency. The core claim is that many organizations fail to realize performance gains not because effort is misallocated or incentives are weak, but because internal organizational mechanisms generate unpriced internal externalities that cause effort to cancel across objectives. These internal conflicts impose a structural ceiling on realized performance that cannot be overcome by optimization alone.

The analysis introduced a minimal framework in which organizational mechanisms (“primitives”) exert signed effects across multiple performance objectives. From these effects, the paper derived measures of magnitude, direction, and conflict and showed how conflict enters realized performance as a multiplicative discount. This formulation separates effort, leverage, and coherence and allows allocative and structural inefficiency to be distinguished analytically.

Applying the framework across several canonical organizational settings— incentive-intensive multitasking environments, governance-heavy organizations, KPI-saturated performance regimes, and multi-dimensional "cheese-grater" reorganizations— the paper demonstrated that structural inefficiency arises endogenously from rational, sequential problem-solving in multi-objective environments. Each setting introduces mechanisms intended to address real deficiencies. Yet because such mechanisms are layered rather than substituted, they expand cross-objective exposure and generate mixed-sign effects that accumulate over time.

A central result of the paper is that structural inefficiency plausibly dominates allocative inefficiency in mature organizations. Stylized business cases show that once internal conflict is widespread among high-effort mechanisms, improvements in allocation, incentives, or analytics yield diminishing returns. Optimization fails not because it is poorly executed, but because the structural yield of effort is intrinsically low. Only reductions in conflict—through simplification, unbundling, or removal of obsolete mechanisms—can unlock substantial gains.

The theory helps explain several persistent organizational puzzles: why performance improvements stall despite high effort; why incentive refinement, governance tightening, and metric proliferation often disappoint; why repeated reorganization produces activity without improvement; and why radical simplification sometimes succeeds where incremental optimization fails. Importantly, the theory does not imply that incentives, governance, metrics, or matrices are inherently flawed. It shows instead that their cumulative interaction, when left unpriced and unmanaged, generates internal conflict that constrains performance.

The paper is intentionally static and theoretical. It does not propose measurement instruments, empirical identification strategies, or organizational interventions. These are deferred to companion work. The contribution here is conceptual: to clarify a structural limit that precedes and conditions optimization.

Future research can extend the framework in several directions. Empirical work can estimate conflict directly by measuring cross-objective effects of organizational mechanisms. Dynamic models can study how conflict accumulates over time and how organizations escape high-conflict traps. Applied work can explore governance mechanisms for pricing or internalizing internal externalities. Together, such extensions would help move from diagnosis to design.

## Appendix A. Formal properties and extensions

This appendix collects formal results that support the arguments in the main text. Proofs are straightforward and included for completeness.

### Bounds on conflict

**Lemma A1.**

For any primitive $i$, conflict satisfies $C_i \in [0,1]$.

**Proof.**

By definition, $M_i = \sum_k \mid b_{ik} \mid \geq 0$ and $\mid D_i \mid = \mid \sum_k b_{ik} \mid \leq \sum_k \mid b_{ik} \mid = M_i$

Hence $0 \leq \mid D_i \mid / M_i \leq 1$, implying $0 \leq C_i \leq 1$.

### Adding objectives weakly increases expected conflict

**Proposition A1.**

Suppose a primitive has effect vector $b_i \in \mathbb{R}^K$. Adding an additional objective $k'$ with effect $b_{i,k'} \neq 0$ weakly increases expected conflict unless $b_{i,k'}$ has the same sign as $D_i$.

**Proof.**

After adding the objective,

$$M_i' = M_i + \mid b_{i,k'} \mid$$

$$D_i' = D_i + b_{i,k'}$$

Conflict decreases only if

$$\frac{\mid D_i' \mid}{M_i'} > \frac{\mid D_i \mid}{M_i},$$

which requires perfect directional alignment. Otherwise, $M_i'$ grows faster than $\mid D_i' \mid$, implying $C_i' \geq C_i$.

**Effort-weighted conflict and aggregate performance**

Define effort-weighted average conflict as:

$$\bar{C}_E = \frac{\sum_i E_i M_i C_i}{\sum_i E_i M_i}$$

**Proposition A2.**
Aggregate realized performance satisfies:

$$R = \left(\sum_i E_i M_i\right)(1 - \bar{C}_E)$$

**Proof.**
Substitute the definition of $R_i^{\text{realized}}$ and collect terms:

$$R = \sum_i E_i M_i - \sum_i E_i M_i C_i = \left(\sum_i E_i M_i\right)\left(1 - \frac{\sum_i E_i M_i C_i}{\sum_i E_i M_i}\right).$$

**Dominance of structural inefficiency**

Let $\alpha \in [0,1]$ represent allocative efficiency, so that

$$\sum_i E_i M_i = \alpha \cdot \max_E \sum_i E_i M_i$$

Then:

$$R = \alpha(1 - \bar{C}_E)R^{\max}.$$

When $\bar{C}_E$ is large, marginal improvements in $\alpha$ have limited effect on $R$, establishing structural inefficiency as the binding constraint.

## Appendix B. Conflict pricing and the internalization of internal externalities

This appendix formalizes a concept that has been implicit throughout the paper: conflict pricing. The core idea is that internal conflict behaves like an unpriced externality inside the firm. When organizational mechanisms generate mixed-sign effects across objectives, they impose costs on other parts of the organization that are not reflected in local decision-making. Conflict pricing introduces a shadow cost that internalizes these effects.

The appendix proceeds in five steps. First, it motivates conflict pricing by analogy with standard externality theory. Second, it introduces a shadow-cost representation of conflict. Third, it shows how conflict pricing alters effort allocation and structural evolution. Fourth, it explains why conflict pricing is rare. Finally, it relates conflict pricing to unbundling and structural cleanup.

### Internal conflict as an internal externality

In standard economic analysis, an externality arises when an action imposes costs or benefits on others that are not priced into the decision-maker's objective function. The result is overuse of activities with negative externalities and underuse of activities with positive ones.

The paper's central claim is that internal conflict is an internal externality. When a primitive $i$ improves some KPIs and worsens others, the negative effects are often borne by different units, time horizons, or stakeholders than the positive ones. Because these cross-effects are not priced, the primitive appears more attractive locally than it is globally.

This logic applies equally to incentives, governance mechanisms, metrics, and organizational dimensions. Each is introduced to improve a specific objective, while its negative cross-effects are diffuse, delayed, or politically difficult to attribute.

### A shadow-cost representation of conflict

To formalize conflict pricing, consider the realized-performance contribution of primitive $i$:

$$R_i^{\text{realized}} = E_i M_i (1 - C_i).$$

Rewrite this as:

$$R_i^{\text{realized}} = E_i M_i - E_i M_i C_i.$$

The second term, $E_i M_i C_i$, represents lost potential due to internal conflict. We interpret this term as an implicit cost imposed by primitive $i$ on the organization.

Define the conflict cost of primitive $i$ as:

$$\kappa_i \equiv E_i M_i C_i.$$

This cost is not monetary in a narrow sense. It represents lost speed, lost quality, coordination burden, frustration, delay, and foregone opportunities caused by cross-objective cancellation.

**Conflict pricing and modified effort allocation**

Suppose the organization could internalize this cost by assigning a shadow price $\lambda > 0$ to conflict. Then the organization's objective becomes:

$$\max_{\{E_i\}} \sum_i [E_i M_i - \lambda E_i M_i C_i]$$

subject to effort constraints.

Equivalently:

$$\max_{\{E_i\}} \sum_i E_i M_i (1 - \lambda C_i).$$

**Interpretation**

- When $\lambda = 0$, conflict is ignored; effort flows toward high-magnitude primitives regardless of conflict.
- When $\lambda > 0$, conflict is penalized; high-conflict primitives become less attractive.

- As $\lambda$ increases, effort is reallocated away from conflicted mechanisms toward more coherent ones, even if their raw magnitude is lower.

This representation makes clear that conflict pricing operates orthogonally to incentives and effort optimization. It does not require changing behavior within primitives; it changes which primitives are emphasized.

**Structural evolution under conflict pricing**

If conflict is unpriced, adding a new dimension increases perceived leverage because it increases magnitude. The organization therefore has a bias toward dimensional expansion.

With conflict pricing, the net benefit of adding a dimension becomes:

$$\Delta \text{Value} = \Delta M_i - \lambda \Delta (M_i C_i).$$

Unless the new dimension is well aligned, the second term dominates for sufficiently large $\lambda$, making dimensional expansion unattractive. Conflict pricing therefore acts as a structural brake.

In this sense, conflict pricing is the formal counterpart of "organizational discipline" or "design coherence," but expressed in analytic rather than normative terms.

**Why conflict pricing is rare in practice**

Despite its conceptual appeal, explicit conflict pricing is rare. The framework helps explain why.

1. **Measurement difficulty**
   Conflict is multi-dimensional and distributed. While magnitude is visible, cancellation is often delayed or contested.
2. **Political asymmetry**
   The benefits of a primitive are typically concentrated and attributable; the costs are diffuse. Pricing conflict creates losers with identifiable sponsors.

3. **Institutional inertia**
   Existing mechanisms embody past priorities. Assigning a shadow cost to conflict implicitly devalues those priorities.
4. **Cognitive framing**
   Organizations are trained to think in terms of optimization, not structural yield. Conflict pricing requires a shift from “doing better” to “doing less but cleaner.”

These obstacles explain why organizations tend to respond to conflict by adding controls or metrics rather than pricing or removing mechanisms—ironically increasing conflict further.

### Conflict pricing versus unbundling

Conflict pricing does not require eliminating mechanisms. It creates incentives to unbundle them.

Suppose a primitive with high magnitude and high conflict can be decomposed into two primitives with lower conflict. Under conflict pricing, such decomposition becomes attractive because it preserves leverage while reducing the shadow cost.

In this sense, conflict pricing provides a formal rationale for organizational simplification strategies often described informally as “clarifying accountability,” “reducing handoffs,” or “removing friction.”

### Relation to allocative inefficiency

Conflict pricing primarily addresses structural inefficiency. It does not solve allocative inefficiency directly. However, by reducing conflict, it increases the structural yield of effort, making allocative improvements more effective.

This reinforces the paper’s central sequencing claim: structural cleanup precedes optimization.

This appendix has shown that internal conflict can be treated as an internal externality and that introducing a shadow price for conflict fundamentally alters both effort allocation and structural evolution. Conflict pricing suppresses cheese-grater dynamics, encourages unbundling, and raises the returns to optimization.

The rarity of conflict pricing in practice helps explain the persistence of structurally inefficient organizations and the recurrent failure of optimization efforts in high-conflict environments.